\documentclass[pra,showpacs,twocolumn,aps,superscriptaddress,a4paper]{revtex4-1}
\usepackage{dcolumn,amssymb,amsmath,amsfonts,graphicx,latexsym,color}
\usepackage{epstopdf}
\usepackage{multirow}
\usepackage{CJK}

\begin{document}

\begin{CJK*}{GBK}{song}

\title{High-momentum tail and universal relations of a Fermi gas near a Raman-dressed Feshbach resonance}

\author{Fang Qin}
\email{qinfang@ustc.edu.cn}
\affiliation{Key Laboratory of Quantum Information, University of Science and Technology of China, Chinese Academy of Sciences, Hefei, Anhui 230026, China}
\affiliation{Synergetic Innovation Center of Quantum Information and Quantum Physics, University of Science and Technology of China, Hefei, Anhui 230026, China}
\author{Jianwen Jie}
\affiliation{Department of Physics, Renmin University of China, Beijing, 100872, China}
\author{Wei Yi}
\email{wyiz@ustc.edu.cn}
\affiliation{Key Laboratory of Quantum Information, University of Science and Technology of China, Chinese Academy of Sciences, Hefei, Anhui 230026, China}
\affiliation{Synergetic Innovation Center of Quantum Information and Quantum Physics, University of Science and Technology of China, Hefei, Anhui 230026, China}
\author{Guang-Can Guo}
\affiliation{Key Laboratory of Quantum Information, University of Science and Technology of China, Chinese Academy of Sciences, Hefei, Anhui 230026, China}
\affiliation{Synergetic Innovation Center of Quantum Information and Quantum Physics, University of Science and Technology of China, Hefei, Anhui 230026, China}

\date{\today}

\begin{abstract}
In a recent proposal [Jie and Zhang, Phys. Rev. A 95, 060701(R) (2017)], it has been shown that center-of-mass-momentum-dependent two-body interactions can be generated and tuned by Raman coupling the closed-channel bound states in a magnetic Feshbach resonance. Here we investigate the universal relations in a three-dimensional Fermi gas near such a laser modulated $s$-wave Feshbach resonance.
Using the operator-product expansion approach, we find that, to fully describe the high-momentum tail of the density distribution up to $q^{-6}$ ($q$ is the relative momentum), four center-of-mass-momentum-dependent parameters are required, which we identify as contacts. These contacts appear in various universal relations connecting microscopic and thermodynamic properties. One contact is related to the variation of energy with respect to the inverse scattering length and determines the leading $q^{-4}$ tail of the high-momentum distribution.
Another vector contact appears in the subleading $q^{-5}$ tail, which is related to the velocity of closed-channel molecules.
The other two contacts emerge in the $q^{-6}$ tail and are respectively related to the variation of energy with respect to the range parameter and to the kinetic energy of closed-channel molecules.
Particularly, we find that the $q^{-5}$ tail and part of the $q^{-6}$ tail of the momentum distribution show anisotropic features. We derive the universal relations and, as a concrete example, estimate the contacts for the zero-temperature superfluid ground state of the system using a mean-field approach.
\end{abstract}

\maketitle

\section{Introduction}\label{1}
Due to the short-range nature of the two-body interactions in dilute atomic gases, thermodynamic properties of degenerate fermions close to scattering resonances are universal, where the system can be described by a handful of physical parameters and are independent of the short-range details of the interaction potentials. In such strong-coupling regimes, universal relations exist among microscopic and thermodynamic quantities, which are connected by a set of key parameters called the contacts. First derived by Tan for a three-dimensional Fermi gas near an $s$-wave Feshbach resonance (FR)~\cite{Tan20081,Tan20082,Tan20083}, contacts and the corresponding universal relations have been experimentally confirmed~\cite{exp_contact1,exp_contact2,exp_contact3} and have been generalized to various situations such as quantum gases in low dimensions~\cite{Zwerger2011,Cui20161,Cui20162,Klumper2017,Castin20121,Castin20122,Valiente2011,Valiente2012,Peng20161,Zhang2017,Zhou2017},
systems with higher or mixed partial-wave scatterings~\cite{Yu2015,Yoshida2015,Yoshida2016,Zhou20161,Zhou20162,Peng20162,Qin2016,Yu2015exp}, bosonic gases~\cite{BraatenBose2011,BraatenBose2014,exp_Jin2012,exp_Zwierlein2017,Vignolo2017,Vignolo2018}, and Fermi gases under synthetic gauge field~\cite{Peng2017,Zhang2018,Jie2018}.

\begin{figure}
\includegraphics[width=7cm]{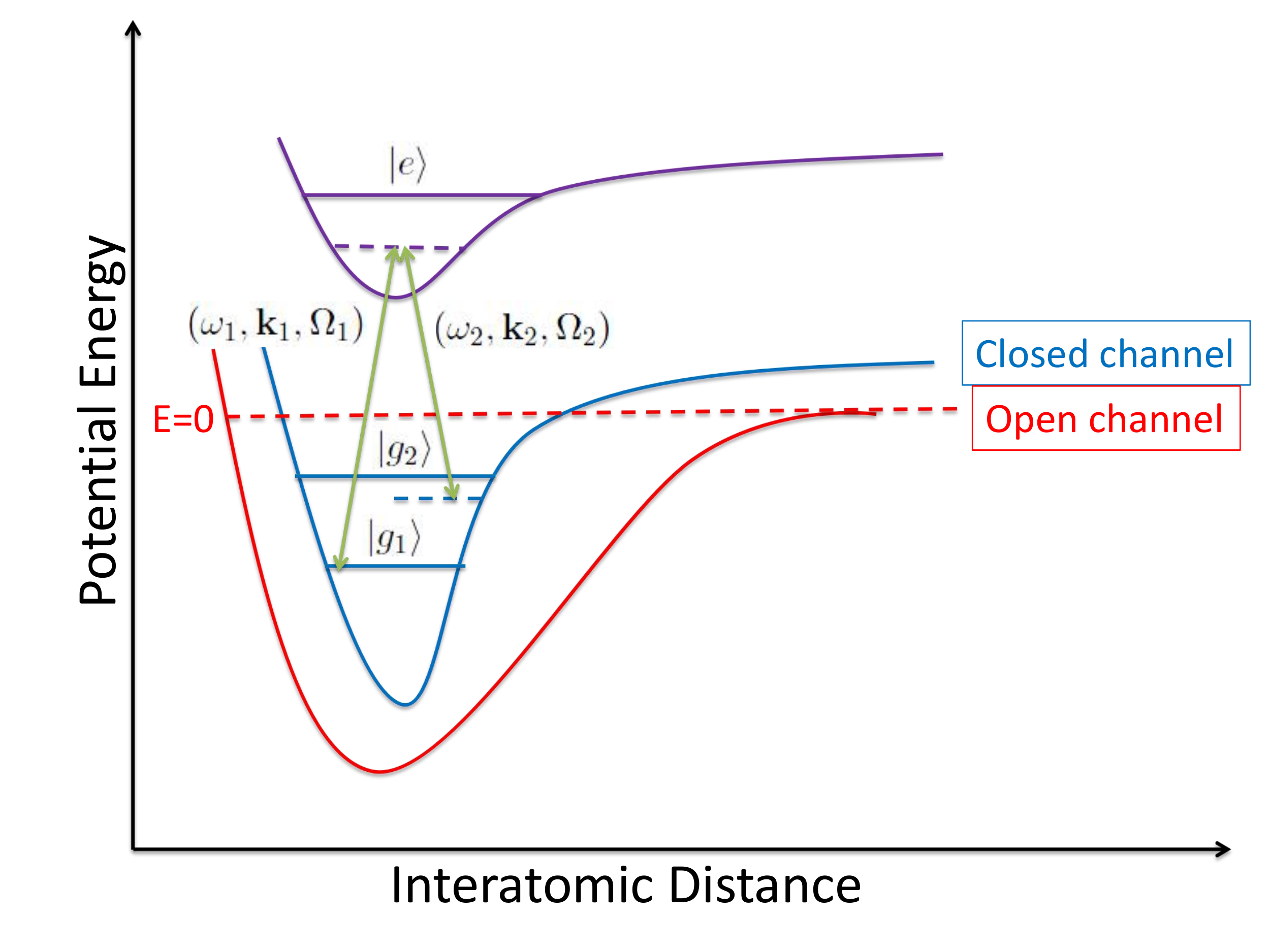}
\caption{(Color online) Level scheme for the Raman-dressed FR, in which the Raman laser beams with frequencies $\omega_1$ and $\omega_2$ propagate along different directions (${\bf k}_{1}\neq{\bf k}_{2}$)~\cite{Jie2017}. The closed-channel bound states are labeled as $|g_1\rangle$, $|g_2\rangle$, and $|e\rangle$.} \label{fig:model}
\end{figure}

Recently, Jie and Zhang proposed an experimental scheme to generate center-of-mass(CoM)-momentum dependent two-body interactions in cold-atomic gases, where the closed-channel bound states in a magnetic FR are Raman coupled~\cite{Jie2017}. As illustrated in Fig.~\ref{fig:model}, the FR is modulated by two counterpropagating optical fields, which are applied to couple two molecular states in the closed channel to an excited molecular state. This leads to a significant Doppler-shifted Stark effect, which causes both the scattering length and range parameter to be strongly dependent on the CoM momentum of the two scattering particles~\cite{Jie2017}. Whereas such a Raman-dressed FR can give rise to a Fulde-Ferrell pairing superfluid in a many-body setting~\cite{He2017}, one also expects that the contacts as well as the universal relations of the system should be significantly modified by the CoM dependence of the interaction.

In this work, we study the high-momentum tail of the density distribution and universal relations of a three-dimensional Fermi gas near a Raman-dressed $s$-wave FR. We adopt the operator-product expansion (OPE) approach~\cite{Wilson,Kadanoff,Braaten20081,Braaten20082,Braaten20083,Platter2016,Yu20171,Yu20172,Yu20173,Qi2016}
and derive the high-momentum distribution. We find that the high-momentum tail can be fully determined, up to $q^{-6}$ ($q$ is the relative momentum of the two-body scattering process), by four parameters which we identify as contacts. We show that one contact is related to the variation of energy with respect to the inverse scattering length, which determines the leading $q^{-4}$ tail of the high-momentum distribution.
Another vector contact shows up in the subleading $q^{-5}$ tail, which is related to the velocity of closed-channel molecules. The other two contacts emerge in the $q^{-6}$ tail and are related to the variation of energy with respect to the range parameter and the kinetic energy of closed-channel molecules, respectively.
Whereas the contact associated with the scattering length is formally similar to Tan's contact in the original context, as both the scattering length and the range parameter are CoM-momentum dependent under the Raman coupling, all four contacts are now CoM-momentum dependent.

We then derive the universal relations and numerically evaluate the contacts for the zero-temperature pairing superfluid state under the Raman-induced CoM-momentum-dependent interactions.
We show that two of the contacts are associated with the mean velocity and the mean kinetic energy of closed-channel molecules, and these may be nonzero when the ground state of the
system is a Fulde-Ferrell-like superfluid. The latter case is expected, at least at the mean-field level, when the two-component gas is exposed to a suitable Raman dressing.

The paper is organized as follows: In Sec.~\ref{2}, we present the two-channel model with Raman-coupled molecular states in the closed channel. We also discuss interaction renormalization in the presence of Raman-coupling fields and calculate the corresponding CoM-momentum-dependent scattering length and range parameter. In Sec.~\ref{3}, we calculate the high-momentum distribution of the system with the OPE approach. In Sec.~\ref{4}, we derive the universal relations such as the adiabatic relations, the pressure relations, the virial theorem, and the energy functional. We numerically evaluate the contacts using a concrete example in Sec.~\ref{5}. Finally, we summarize in Sec.~\ref{6}.

\section{Two-channel model and interaction renormalization}\label{2}
In the presence of the Raman lasers, the local Lagrangian density (at coordinate ${\bf R}$) is given by ${\cal L} = {\cal L}_{{\rm A}} + {\cal L}_{{\rm M}} + {\cal L}_{{\rm AM}}$, where~\cite{He2017}
\begin{align}
{\cal L}_{{\rm A}}& = \sum_{\sigma=\uparrow,\downarrow}\psi_{\sigma}^{\dagger}\left(i\partial_{t}+\frac{\nabla^{2}_{\bf R}}{2m}\right)\psi_{\sigma}, \label{eq:LA}\\
{\cal L}_{{\rm M}} & =\sum_{l=1,2} \varphi_{l}^{\dagger}\left(i\partial_{t}+\frac{\nabla^{2}_{\bf R}}{4m}-E_l\right)\varphi_{l} \nonumber \\
&+ \varphi_{{\rm e}}^{\dagger}\left(i\partial_{t}+\frac{\nabla^{2}_{\bf R}}{4m}-E_{{\rm e}}+i\frac{\gamma_{{\rm e}}}{2}\right)\varphi_{{\rm e}} \nonumber \\
&- \sum_{l=1,2}\left(\frac{\Omega_{l}}{2}\varphi_{{\rm e}}^{\dagger}\varphi_{l}e^{i\theta_{l}}
+ \frac{\Omega_{l}^{*}}{2}\varphi_{l}^{\dagger}\varphi_{{\rm e}}e^{-i\theta_{l}}\right), \label{eq:LM}\\
{\cal L}_{{\rm AM}}&  = -g_{0}\left(\varphi_{1}^{\dagger}\psi_{\downarrow}\psi_{\uparrow} + \psi_{\uparrow}^{\dagger}\psi_{\downarrow}^{\dagger}\varphi_{1}\right). \label{eq:LAM}
\end{align}
Here, $\psi_{\sigma}$ ($\sigma=\uparrow,\downarrow$) are the field operators for the open-channel fermions. $\varphi_{l}$ ($l=1,2$) and $\varphi_{e}$ are the field operators for the closed-channel molecules in the states $|g_l\rangle$ ($l=1,2$) and $|e\rangle$, respectively. $m$ is the atomic mass. The closed-channel molecules in the state $|g_l\rangle$ are coupled to the excited state $|e\rangle$ by the Raman laser
with the strength $\Omega_{l}/2$ and the phase $\theta_{l}({\bf R},t)={\bf k}_{l}\cdot{\bf R}-\omega_{l}t$. ${\bf k}_{l}$ and $\omega_{l}$ are the wave vector and the frequency of the corresponding optical field. The energies of the molecular states with respect to the open-channel threshold are given by $E_l$ and $E_e$, respectively. In the following, we will denote the bare molecular detuning $E_1=\nu_0$ following the common practice. $\gamma_{{\rm e}}$ denotes the spontaneous decay rate of the excited molecular state $|e\rangle$. The coupling between the open and the closed channel is given by $g_{0}$.
Note that we have neglected the background fermion-fermion interactions in the open channel, since it is not important close to resonance~\cite{renormalization2004}. We have taken the natural units $\hbar=k_B=1$ throughout the paper.

Following the practice in Ref.~\cite{He2017}, we remove the phase factor $e^{\pm i\theta_{l}}$
($l=1,2$) in the last term of ${\cal L}_{{\rm M}}$, and introduce two new molecular fields: $\phi_{{\rm e}}({\bf R})=\varphi_{{\rm e}}({\bf R})e^{-i\theta_{1}}$
and $\phi_{2}({\bf R})=\varphi_{2}({\bf R})e^{-i(\theta_{1}-\theta_{2})}$. We can then write the molecular part in the momentum space: ${\cal L}_{{\rm M}}=\Phi^{\dagger}_{\bf Q}M(q_{0},{\bf Q})\Phi_{\bf Q}$,
where $\Phi_{\bf Q}=(b_{{\bf Q},1},\ b_{{\bf Q},2},\ b_{{\bf Q},{\rm e}})^{{\rm T}}$ and
the inverse propagator matrix is given by
\begin{eqnarray}
M(q_{0},{\bf Q})=\left(\begin{array}{ccc}
I_{1}(q_{0},{\bf Q}) & 0 & -\Omega_{1}^{*}/2\\
0 & I_{2}(q_{0},{\bf Q}) & -\Omega_{2}^{*}/2\\
-\Omega_{1}/2 & -\Omega_{2}/2 & I_{{\rm e}}(q_{0},{\bf Q})
\end{array}\right)
\end{eqnarray}
with
\begin{eqnarray}
 &  & I_{1}(q_{0},{\bf Q}) = q_{0}-\frac{{\bf Q}^{2}}{4m} - \nu_{0}, \label{eq:I1}\\
 &  & I_{2}(q_{0},{\bf Q}) = q_{0}-\frac{{\bf Q}^{2}}{4m} - \Delta_{2}({\bf Q}), \label{eq:I2} \\
 &  & I_{{\rm e}}(q_{0},{\bf Q}) = q_{0}-\frac{{\bf Q}^{2}}{4m} - \Delta_{e}({\bf Q}) + i\frac{\gamma_{{\rm e}}}{2}, \label{eq:Ie} \\
 &  & \Delta_{2}({\bf Q}) = \frac{\mathbf{Q}\cdot({\bf k}_{1}-{\bf k}_{2})}{2m} + \frac{({\bf k}_{1}-{\bf k}_{2})^{2}}{4m} + \delta_2, \label{eq:D2} \\
 &  & \Delta_{e}({\bf Q}) = \frac{{\bf Q}\cdot{\bf k}_{1}}{2m} + \frac{{\bf k}_{1}^{2}}{4m} + \delta_{{\rm e}}. \label{eq:De}
\end{eqnarray}
Here, $\delta_2 = E_{2} - (\omega_{1}-\omega_{2})$ is the two-photon detuning, $\delta_{{\rm e}} = E_{{\rm e}} - \omega_{1}$ is the one-photon
detuning, ${\bf Q}$ is the CoM momentum, and $q_{0}$ is the total incoming energy. In the following discussions, we assume that $\varphi_2$ is far detuned from the open channel and therefore not resonant.

\begin{figure}
\includegraphics[width=7cm]{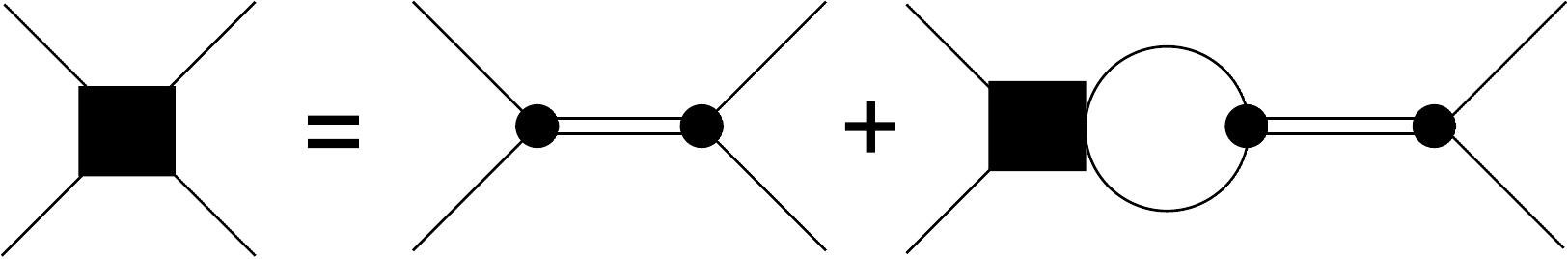}
\caption{Diagram for calculating the $T$ matrix. The single lines denote the bare atom propagators $G_{0}$ and the double lines denote the bare molecular propagators $D_{1}$. The black solid square represents the $T$ matrix: $-iT$. The black dots represent the interaction vertices: $-i g_{0}$.} \label{fig:T}
\end{figure}

As illustrated in Fig.~\ref{fig:T}, in the presence of the Raman coupling, the two-body $T$ matrix is given by
\begin{eqnarray}\label{eq:T}
-iT(q_0,{\bf Q}) = \frac{(-ig_{0})^{2}D_{1}(q_0,{\bf Q})}{ 1 - (-ig_{0})^{2}D_{1}(q_0,{\bf Q})\Pi(q_0,{\bf Q})},
\end{eqnarray}
where the bare molecule propagator is
\begin{eqnarray}
D_{1}(q_0,{\bf Q}) = \frac{i}{q_{0} - {\bf Q}^{2}/(4m) - \nu_0 - \Sigma_1(q_0,{\bf Q}) + i0^+} \nonumber\\
\end{eqnarray} with
the self-energy~\cite{He2017}
\begin{eqnarray}
\Sigma_{1}(q_{0},{\bf Q})=
\frac{|\Omega_{1}|^{2}}{4}\left[I_{{\rm e}}(q_{0},{\bf Q})-\frac{|\Omega_{2}|^{2}}{4I_{2}(q_{0},{\bf Q})}\right]^{-1}, \label{eq: StarkShift}
\end{eqnarray}
which is essentially the Stark shift of the state $|g_1\rangle$ under the Raman laser. The polarization bubble is
\begin{eqnarray}
\Pi(q_0,{\bf Q})
= \int \frac{d^{3}{\bf p}}{(2\pi)^3} \frac{i}{q_{0} - {\bf Q}^{2}/(4m) - p^2/m + i0^+}. \nonumber\\
\end{eqnarray}

We consider the scattering between two fermions with momenta ${\bf Q}/2+{\bf k}$ and ${\bf Q}/2-{\bf k}$ and with a total energy $q_0=Q^2/(4m)+k^2/m$. In the absence of the Raman coupling, the bare atom-molecule coupling $g_0$ and the bare detuning $\nu_0$ can be renormalized as~\cite{renormalization2004}
\begin{align}
& g_0= g, \label{renormalization2a0}\\
& \nu_0 = \nu +\frac{m g_{0}^2\Lambda}{2\pi^2}, \label{renormalization2b0}
\end{align}
where $\Lambda$ is the ultraviolet momentum cutoff, and the renormalized parameters are
\begin{align}
& g^2 = -\frac{4\pi\nu a}{m}, \label{renormalization2a}\\
& \nu = (\delta\mu)(B - B_0). \label{renormalization2b}
\end{align}
Here, $a$ is the scattering length for fermions in the open channel, $B_0$ is the FR point, and $\delta\mu$ is the difference in magnetic moments between the open and closed channels.

In the presence of the Raman fields, we match the scattering amplitude with the $T$ matrix,
\begin{eqnarray}\label{eq:f}
f({\bf k},{\bf Q})
&=& -\frac{m}{4\pi}T({\bf k},{\bf Q}) \nonumber\\
&=& \frac{1}{ - \frac{4\pi(-\nu_0)}{m g_{0}^{2}} - \frac{2\Lambda}{\pi} - \frac{4\pi}{m g_{0}^{2}}\left[\frac{k^2}{m} - \Sigma_1({\bf k},{\bf Q})\right] - ik} \nonumber\\
&\simeq& \frac{1}{ - 1/a - k^2R_s - ik},
\end{eqnarray}
from which, we can relate the scattering length $a$ and the range parameter $R_s$ with the bare parameters,
\begin{align}
\frac{1}{a} &= \frac{4\pi}{m g_{0}^{2}}\left[- \nu_0 + \frac{|\Omega_{1}|^{2}/4}{ \Delta_{{\rm e}}({\bf Q}) - i\frac{\gamma_{{\rm e}}}{2} - \frac{|\Omega_{2}|^{2}}{4\Delta_{2}({\bf Q})} } \right] + \frac{2\Lambda}{\pi}, \\
R_s &= \frac{4\pi}{m^{2} g_{0}^{2}}\left\{ 1 + \frac{\frac{|\Omega_{1}|^{2}}{4}\left[1 + \frac{|\Omega_{2}|^{2}}{4\Delta_{2}^{2}({\bf Q})}\right]}{\left[\Delta_{{\rm e}}({\bf Q}) - i\frac{\gamma_{\rm e}}{2} - \frac{|\Omega_{2}|^{2}}{4\Delta_{2}({\bf Q})} \right]^2 }\right\}.
\end{align}
Importantly, both the scattering length and the range parameter are now CoM-momentum dependent.

\section{High-momentum distribution}\label{3}
In this section, we derive the high-momentum distribution for fermions with Raman-dressed FR using the quantum field method of OPE~\cite{Zwerger2011,Cui20161,Cui20162,Wilson,Kadanoff,Braaten20081,Braaten20082,Braaten20083,Platter2016,Yu20171,Yu20172,Yu20173,Qi2016,Zhang2018}.

The expression of the momentum distribution for the spin-$\sigma$ species is given as~\cite{Braaten20082}
\begin{align}\label{eq:rho}
n_{\sigma}({\bf q}) =
\int \frac{d^3{\bf R}}{V} \int d^3{\bf r} e^{-i{\bf q}\cdot{\bf r}} \left\langle \psi_{\sigma}^{\dag}({\bf R}+\frac{{\bf r}}{2}) \psi_{\sigma}({\bf R}-\frac{{\bf r}}{2})  \right\rangle. \nonumber\\
\end{align}
where $V$ is the system volume.
As the large-$q$ behavior of $n_{\sigma}({\bf q})$ is essentially determined by the one-body density matrix $\langle \psi_{\sigma}^{\dag}({\bf R}+\frac{{\bf r}}{2}) \psi_{\sigma}({\bf R}-\frac{{\bf r}}{2}) \rangle$
at short distances $r\ll 1/q$, we have the expansion
\begin{equation}\label{ope}
\psi_{\sigma}^{\dag}({\bf R}+\frac{{\bf r}}{2}) \psi_{\sigma}({\bf R}-\frac{{\bf r}}{2}) = \sum_n C_n({\bf r}) {\cal O}_n({\bf R}),
\end{equation}
where the local operator ${\cal O}_n({\bf R})$ can be constructed by quantum fields and their derivatives, and non-analytic dependence of the coefficients $C_n({\bf r})$ on ${\bf r}$ gives rise to the high-momentum tail of $n_{\sigma}({\bf q})$.

To determine $C_n({\bf r})$, we calculate the matrix elements of the operators on both sides of Eq.~(\ref{ope}) for an incoming state and an outgoing state in a two-body scattering process. We then identify the corresponding local operators ${\cal O}_n({\bf R})$ by matching the left- and right-hand sides of Eq.~(\ref{ope}). We consider the incoming state $|I\rangle=|{\bf Q}/2 + {\bf k},\uparrow; {\bf Q}/2 - {\bf k},\downarrow\rangle$, where the two fermions have momentum ${\bf Q}/2 + {\bf k}$ and ${\bf Q}/2 - \mathbf{k}$, respectively. Similarly, the outgoing state can be written as $|O\rangle=|{\bf Q}/2 + {\bf k}',\uparrow; {\bf Q}/2 - {\bf k}',\downarrow\rangle$.

\begin{figure}
\includegraphics[width=5cm]{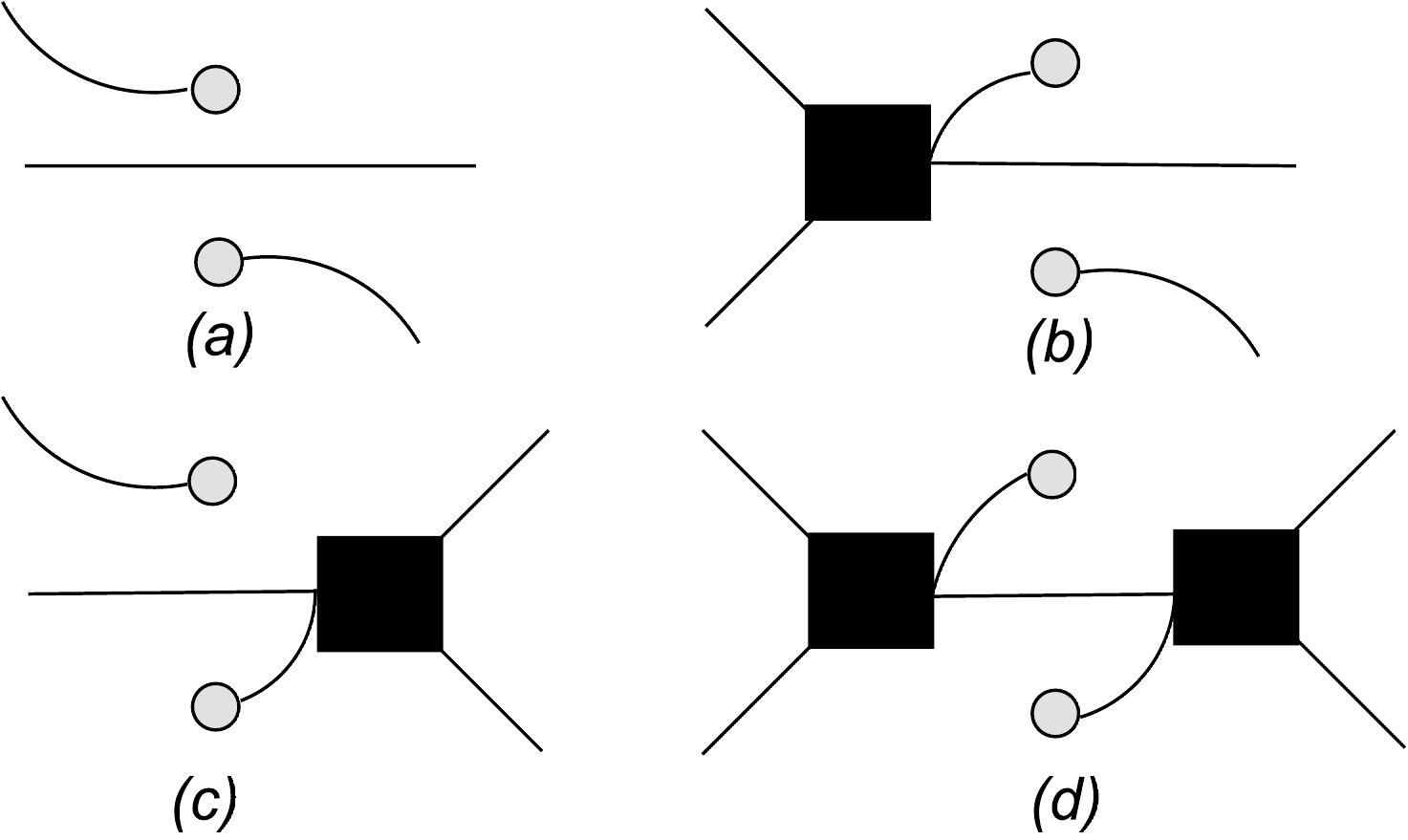}
\caption{Diagrams for matrix elements of the
operator $\psi_\sigma^\dagger({\bf R} + \frac{1}{2}{\bf r})\psi_\sigma({\bf R} -\frac{1}{2}{\bf r})$. The gray dots represent the operators. \label{fig:OneBodyOperator}}
\end{figure}

\begin{figure}
\includegraphics[width=5cm]{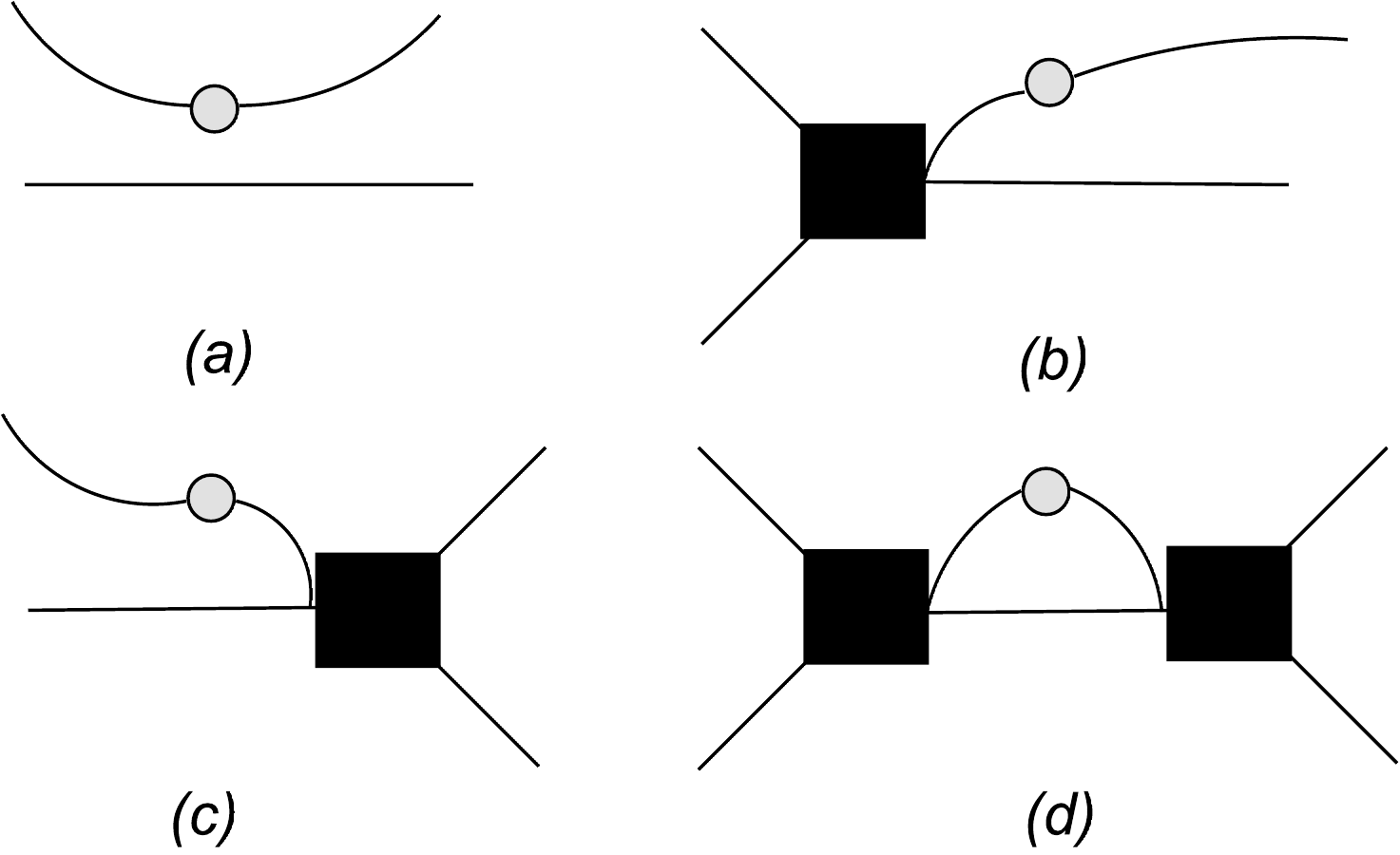}
\caption{Diagrams for the matrix elements of the
one-atom local operator $\psi_\sigma^\dagger({\bf R})\psi_\sigma({\bf R})$ and its derivatives. The gray dots represent the operators. \label{fig:OneBodyOperatorR}}
\end{figure}

The left-hand side of Eq.~(\ref{ope}) can produce four types of diagrams, which are shown in Figs.~\ref{fig:OneBodyOperator}(a)-\ref{fig:OneBodyOperator}(d).
The diagrams in Figs.~\ref{fig:OneBodyOperator}(a)-\ref{fig:OneBodyOperator}(c) produce analytic functions of ${\bf r}$, which can be matched by matrix elements of one-atom local operator $\psi_\sigma^\dagger({\bf R})\psi_\sigma({\bf R})$ and its derivatives, as shown by Figs.~\ref{fig:OneBodyOperatorR}(a)-\ref{fig:OneBodyOperatorR}(c). The only nonanalytic terms come from the diagram shown in Fig.~\ref{fig:OneBodyOperator}(d), which also contains analytical terms given by Fig.~\ref{fig:OneBodyOperatorR}(d). More explicitly, we can evaluate the diagram in Fig.~\ref{fig:OneBodyOperator}(d) as
\begin{widetext}
\begin{align}
&~~\left\langle O\left| \psi_\sigma^{\dag}({\bf R}+\frac{{\bf r}}{2}) \psi_\sigma({\bf R}-\frac{{\bf r}}{2}) \right|I\right\rangle_{d} \nonumber\\
&= \int \frac{d^{3}{\bf p} dp_0}{(2\pi)^4} \frac{[-iT({\bf k},{\bf Q})]^2 i^3 e^{-i({\bf Q}/2+{\bf p})\cdot{\bf r}}}{[p_0 - ({\bf Q}/2-{\bf p})^2/(2m) + i0^+][q_0 - p_0 - ({\bf Q}/2+{\bf p})^2/(2m) + i0^+]^2} \nonumber \\
&= \frac{im^2T^2({\bf k},{\bf Q})}{8\pi} \left\{\frac{1}{k} + i\left[1 - \frac{Q}{2k}P_{1}(\hat{{\bf Q}}\cdot\hat{{\bf r}})\right]r - \left[\frac{k}{2} + \frac{Q^2}{24k} - \frac{Q}{2}P_{1}(\hat{{\bf Q}}\cdot\hat{{\bf r}}) + \frac{Q^2}{12k} P_{2}(\hat{{\bf Q}}\cdot\hat{{\bf r}}) \right]r^2 \right. \nonumber\\
&\left.~~  - i\left[\frac{k^2}{6} + \frac{Q^2}{24} + \left(-\frac{kQ}{4} - \frac{Q^3}{80k}\right) P_{1}(\hat{{\bf Q}}\cdot\hat{{\bf r}}) + \frac{Q^2}{12}P_{2}(\hat{{\bf Q}}\cdot\hat{{\bf r}}) - \frac{Q^3}{120k} P_{3}(\hat{{\bf Q}}\cdot\hat{{\bf r}}) \right]r^3 + {\cal O}(r^4) + \cdot\cdot\cdot \right\}, \label{eq:OneBodyOperator}
\end{align}
\end{widetext}
where we use the Rayleigh expansion~\cite{Fourier3D}
\begin{align}
e^{i({\bf Q}\cdot{\bf r})/2} = \sum_{l=0}^{\infty} (2l+1)i^{l}j_{l}(Qr/2) P_{l}(\hat{{\bf Q}}\cdot\hat{{\bf r}}).
\end{align}
Here, $j_{l}(x)$ and $P_{l}(\hat{{\bf Q}}\cdot\hat{{\bf r}})$ are the spherical Bessel function and the Legendre polynomial, respectively.

\begin{figure}
\includegraphics[width=7cm]{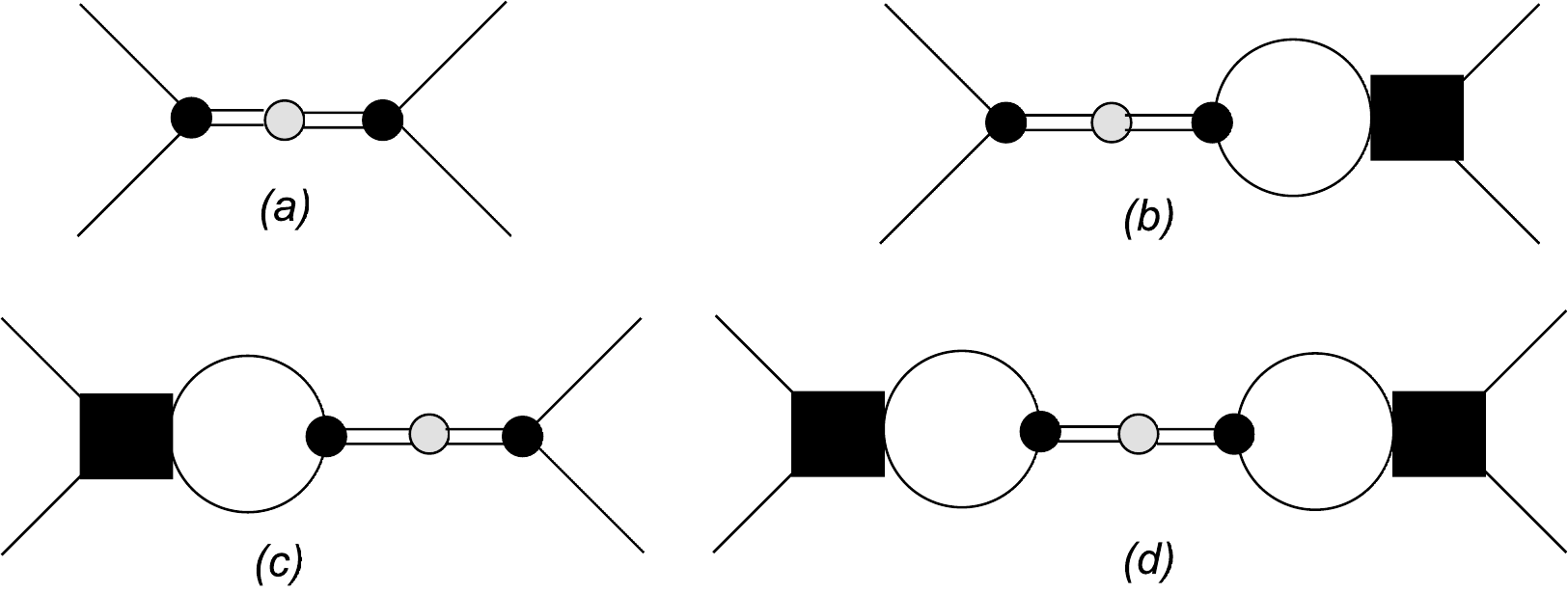}
\caption{Diagrams for matrix elements of the
molecule local operator $\varphi_{1}^\dagger({\bf R}) \varphi_{1}({\bf R})$. The gray dots represent the operators. The black dots represent the interaction vertices. \label{fig:TwoBodyOperatorR}}
\end{figure}

In order to match the non-analytic terms on the right-hand side of Eq.~(\ref{eq:OneBodyOperator}), we calculate the expectation values of the
molecule local operator $\varphi_{1}^\dagger({\bf R}) \varphi_{1}({\bf R})$ according to Fig.~\ref{fig:TwoBodyOperatorR}, which yields
\begin{eqnarray}
&&\langle O| \varphi_{1}^{\dag}({\bf R}) \varphi_{1}({\bf R}) |I\rangle_{a} \nonumber\\
 &=& D_{1}^{2}({\bf k},{\bf Q})(-ig_0)^2, \\
&&\langle O| \varphi_{1}^{\dag}({\bf R}) \varphi_{1}({\bf R}) |I\rangle_{b} \nonumber\\
 &=& D_{1}^{2}({\bf k},{\bf Q})(-ig_0)^2[-iT({\bf k},{\bf Q})]\Pi({\bf k},{\bf Q}), \\
&&\langle O| \varphi_{1}^{\dag}({\bf R}) \varphi_{1}({\bf R}) |I\rangle_{c} \nonumber\\
 &=& D_{1}^{2}({\bf k},{\bf Q})(-ig_0)^2[-iT({\bf k},{\bf Q})]\Pi({\bf k},{\bf Q}), \\
&&\langle O| \varphi_{1}^{\dag}({\bf R}) \varphi_{1}({\bf R}) |I\rangle_{d} \nonumber\\
 &=& D_{1}^{2}({\bf k},{\bf Q})(-ig_0)^2[-iT({\bf k},{\bf Q})\Pi({\bf k},{\bf Q})]^2.
\end{eqnarray}
Therefore, we have
\begin{eqnarray}\label{eq:TwoBodyOperatorR1}
&&\langle O| \varphi_{1}^{\dag}({\bf R}) \varphi_{1}({\bf R}) |I\rangle \nonumber\\
&=& \langle O| \varphi_{1}^{\dag}({\bf R}) \varphi_{1}({\bf R}) |I\rangle_{a} + \langle O| \varphi_{1}^{\dag}({\bf R}) \varphi_{1}({\bf R}) |I\rangle_{b} \nonumber\\
&& + \langle O| \varphi_{1}^{\dag}({\bf R}) \varphi_{1}({\bf R}) |I\rangle_{c} + \langle O| \varphi_{1}^{\dag}({\bf R}) \varphi_{1}({\bf R}) |I\rangle_{d} \nonumber\\
&=& D_{1}^{2}({\bf k},{\bf Q})(-ig_0)^2[1-iT({\bf k},{\bf Q})\Pi({\bf k},{\bf Q})]^2.
\end{eqnarray}

Substituting Eq.~(\ref{eq:T}) into (\ref{eq:TwoBodyOperatorR1}), we have
\begin{eqnarray}\label{eq:TwoBodyOperatorR2}
\langle O| \varphi_{1}^{\dag}({\bf R}) \varphi_{1}({\bf R}) |I\rangle = \frac{T^2({\bf k},{\bf Q})}{g_{0}^{2}}.
\end{eqnarray}
Finally, we get
\begin{align}
&\left\langle O\left| \varphi_{1}^\dagger({\bf R})\left(i\partial_t+\frac{\nabla^2_{{\bf R}}}{4m}\right) \varphi_{1}({\bf R}) \right|I\right\rangle = \frac{k^2}{m}\frac{T^2({\bf k},{\bf Q})}{g_{0}^{2}}, \label{eq:TwoBodyOperatorR3}\\
&\left\langle O\left| \varphi_{1}^\dagger({\bf R})\left(-i\nabla_{{\bf R}}\right) \varphi_{1}({\bf R}) \right|I\right\rangle = {\bf Q}\frac{T^2({\bf k},{\bf Q})}{g_{0}^{2}}, \label{eq:TwoBodyOperatorR4}\\
&\left\langle O\left| \varphi_{1}^\dagger({\bf R})\left(-\frac{\nabla^2_{{\bf R}}}{4m}\right) \varphi_{1}({\bf R}) \right|I\right\rangle = \frac{Q^2}{4m}\frac{T^2({\bf k},{\bf Q})}{g_{0}^{2}}. \label{eq:TwoBodyOperatorR5}
\end{align}

Matching Eq.~(\ref{eq:OneBodyOperator}) with Eqs.~(\ref{eq:TwoBodyOperatorR2})-(\ref{eq:TwoBodyOperatorR5}) and using the Fourier transform given by Eq.~(\ref{eq:rho}), we get the momentum distribution $n_{\sigma}({\bf q})$ of the spin-$\sigma$ species in the large-$q$ limit ($n^{1/3}\ll q\ll 1/r_0$, with $n$ the number density and $r_0$ the interaction range),
\begin{align}\label{eq:tail}
n_{\sigma}({\bf q}) &= \frac{C_a}{Vq^{4}} + \frac{2\hat{{\bf q}}\cdot{\bf C}_{Q1}}{Vq^{5}} \nonumber\\
&~~ + \frac{2[C_{R} - C_{Q2} + 6C_{Q2}(\hat{{\bf q}}\cdot\hat{{\bf Q}})^2]}{Vq^{6}},
\end{align}
where the corresponding contacts are defined as
\begin{align}
&C_{a} \equiv m^2g_{0}^2 \int d^{3}{\bf R} \langle \varphi_{1}^\dagger({\bf R}) \varphi_{1}({\bf R}) \rangle, \label{ca}\\
&C_{R} \equiv m^3 g_{0}^2 \int d^{3}{\bf R} \left\langle \varphi_{1}^\dagger({\bf R})\left(i\partial_t+\frac{\nabla^2_{{\bf R}}}{4m}\right) \varphi_{1}({\bf R}) \right\rangle, \label{cR}\\
&{\bf C}_{Q1} \equiv m^2 g_{0}^2 \int d^{3}{\bf R} \left\langle \varphi_{1}^\dagger({\bf R})\left(-i\nabla_{{\bf R}}\right) \varphi_{1}({\bf R}) \right\rangle, \label{cQ1} \\
&C_{Q2} \equiv m^3 g_{0}^2 \int d^{3}{\bf R} \left\langle \varphi_{1}^\dagger({\bf R})\left(-\frac{\nabla^2_{{\bf R}}}{4m}\right) \varphi_{1}({\bf R}) \right\rangle. \label{cQ2}
\end{align}
Note that ${\bf C}_{Q1}$ is a vector quantity.
Particularly, we find that the $q^{-5}$ tail and part of the $q^{-6}$ tail of the momentum distribution given by Eq.~(\ref{eq:tail}) show anisotropic behaviors which are related to ${\bf C}_{Q1}$ and $C_{Q2}$.

Whereas all four contacts are now dependent on the CoM momentum of the colliding atoms, as we will show below, $C_{a}$ and $C_{R}$ are associated, through the adiabatic relations, to the scattering length and the range parameter of the $s$-wave interaction potential.
The last two contacts ${\bf C}_{Q1}$ and $C_{Q2}$, which do not enter the adiabatic relations above, can be related to the velocity and the kinetic energy of the closed-channel molecules, respectively.
In previous studies, it has been shown that contacts of a similar nature to ${\bf C}_{Q1}$ and $C_{Q2}$ can exist for systems with either $s$-wave~\cite{Castin20121,Castin20122,Peng20162,Jie2018,Platter2016} or $p$-wave~\cite{Cui20162} interactions. For ${\bf C}_{Q1}$ and $C_{Q2}$ to take finite values, however, the system should be either non-equilibrium or in an equilibrium state characterized by a finite CoM momentum.
As we will illustrate later, for a typical ground state of our system, ${\bf C}_{Q1}$ and $C_{Q2}$ acquire finite values as soon as the Raman coupling is switched on.

\section{Universal relations}\label{4}
In this section, we derive the universal relations of our system, in which the contacts defined above serve as key parameters for various thermodynamic properties.

\subsection{Adiabatic relations}\label{4.1}
In order to derive the adiabatic relations, we invoke the Feynman-Hellmann theorem, and examine the derivatives of the total energy $E$ with respect to the two-body parameters
\begin{align}
& \frac{\partial E}{\partial a^{-1}} = -\int d^{3}{\bf R} \left\langle \frac{\partial {\cal L}}{\partial a^{-1}} \right\rangle, \\
& \frac{\partial E}{\partial R_s} =  -\int d^{3}{\bf R} \left\langle \frac{\partial {\cal L}}{\partial R_s} \right\rangle.
\end{align}
Applying the renormalization relations and the equation of motion for $\varphi_{1}({\bf R})$, we have
\begin{align}
\left\langle \frac{\partial {\cal L}}{\partial a^{-1}} \right\rangle
&= \left\langle \frac{\partial {\cal L}}{\partial \nu_{0}} \right\rangle \frac{\partial \nu_{0}}{\partial a^{-1}} = \frac{mg_{0}^2}{4\pi} \langle \varphi_{1}^\dagger({\bf R}) \varphi_{1}({\bf R}) \rangle, \label{ca0}\\
\left\langle \frac{\partial {\cal L}}{\partial R_s} \right\rangle
&= \left\langle \frac{\partial {\cal L}}{\partial g_{0}} \right\rangle \frac{\partial g_{0}}{\partial R_s} + \left\langle \frac{\partial {\cal L}}{\partial \nu_{0}} \right\rangle \frac{\partial \nu_{0}}{\partial g_0} \frac{\partial g_{0}}{\partial R_s} \nonumber\\
&= \frac{m g_{0}^{2}}{4\pi} \left\langle \varphi_{1}^\dagger({\bf R})\left(i\partial_t+\frac{\nabla^2_{{\bf R}}}{4m}\right) \varphi_{1}({\bf R}) \right\rangle. \label{cR0}
\end{align}
We then have the adiabatic relations
\begin{align}
& \frac{\partial E}{\partial a^{-1}} = -\frac{C_{a}}{4\pi m}, \label{eq:caE}\\
& \frac{\partial E}{\partial R_s} =  -\frac{C_{R}}{4\pi m},\label{eq:cRE}
\end{align}
where $C_a$ and $C_R$ are related to the energy derivative with respect to the inverse scattering length and the range parameter, respectively.

\subsection{Pressure relation}\label{4.2}
For a uniform gas, the pressure relation can be derived following the expression of the Helmholtz free-energy density ${\cal F}=F/V$, which can be expressed in terms of a dimensionless function $f$~\cite{Tan20083,Zwerger2011,Cui20161,Cui20162,Yu2015,Platter2016,Zhang2009},
\begin{align}
{\cal F}(T, n, a, R_s)
= \frac{1}{6\pi^2 m}k_{F}^{5} f\left(2m\frac{T}{k_{F}^{2}}, \frac{n}{k_{F}^{3}}, a k_{F}, R_s k_{F} \right), \label{eq:F1}
\end{align}
where $T$ is the system temperature and $k_F$ is the Fermi wave vector.

Equation~(\ref{eq:F1}) implies the scaling behavior of the Helmholtz free-energy density as follows
\begin{align}
\tilde{\lambda}^{5} {\cal F}(T, n, a, R_s) = {\cal F}\left(\tilde{\lambda}^{2}T, \tilde{\lambda}^{3}n, \tilde{\lambda}^{-1}a, \tilde{\lambda}^{-1}R_s \right), \label{eq:F2}
\end{align} for a dimensionless and arbitrary parameter $\tilde{\lambda}$.

Taking the derivative of Eq.~(\ref{eq:F2}) with respect to $\tilde{\lambda}$ at $\tilde{\lambda}=1$ yields
\begin{align}
5{\cal F}
=\left(2T \frac{\partial}{\partial T} + 3n\frac{\partial}{\partial n} - a\frac{\partial}{\partial a} - R_s\frac{\partial}{\partial R_s}\right){\cal F}. \label{eq:F3}
\end{align}

Substituting the thermodynamical relations ${\cal F}=n\mu-{\cal P}$, ${\cal E}={\cal F}+T{\cal S}$, and ${\cal S}=-\partial {\cal F}/\partial T$ into Eq.~(\ref{eq:F3}), together with the adiabatic relations (\ref{eq:caE}) and (\ref{eq:cRE}), we can get the pressure relation as
\begin{equation}
{\cal P} = \frac{2}{3}{\cal E} + \frac{C_a}{12\pi maV} - \frac{R_sC_R}{12\pi mV}, \label{eq:pressure}
\end{equation}
where $\mu$ is the chemical potential, $\cal S$ is the entropy density,
$\cal P$ is the pressure density, and $\cal E$ is the energy density.

\subsection{Virial theorem}\label{4.3}
For an atomic gas in a harmonic potential $V_T=\sum_j m\omega^2 {\bf r}_j^2/2$, the Helmholtz free energy can be expressed in terms of a dimensionless function $\tilde{f}$~\cite{Tan20083,Zwerger2011,Cui20161,Cui20162,Yu2015,Platter2016,Zhang2009},
\begin{align}
F( T, \omega, a, R_s, N )
= \omega \tilde{f}( T/\omega, \omega/\omega, a/a_{ho}, R_s/a_{ho}, N ), \label{eq:FV1}
\end{align}
where $N$ is the particle number and $a_{ho} = \sqrt{2/(m\omega)}$ is the harmonic-oscillator length.

With Eq.~(\ref{eq:FV1}), we can get the scaling law
\begin{align}
\tilde{\lambda} F( T, \omega, a, R_s, N )
= F( \tilde{\lambda}T, \tilde{\lambda}\omega, \tilde{\lambda}^{-1/2}a, \tilde{\lambda}^{-1/2}R_{s}, N ), \label{eq:FV2}
\end{align}
where $\tilde{\lambda}$ is a dimensionless and arbitrary parameter. The derivative of Eq.~(\ref{eq:FV2}) with respect to $\tilde{\lambda}$ at $\tilde{\lambda}=1$ yields
\begin{align}
F=\left(T \frac{\partial}{\partial T} + \omega \frac{\partial}{\partial \omega} - \frac{1}{2}a \frac{\partial}{\partial a} - \frac{1}{2}R_s\frac{\partial}{\partial R_s}\right)F. \label{eq:FV3}
\end{align}
With the Legendre transformation in thermodynamics, one has $E=F+TS$, where the entropy is given by $S=-\partial F/\partial T$. Substituting $E=F+TS$ and $S=-\partial F/\partial T$ into Eq.~(\ref{eq:FV3}), one gets
\begin{align}
E=\left(\omega \frac{\partial}{\partial \omega} - \frac{1}{2}a \frac{\partial}{\partial a} - \frac{1}{2}R_s\frac{\partial}{\partial R_s}\right)E,
\end{align}
which, together with the Feynman-Hellmann theorem and the adiabatic relations (\ref{eq:caE}) and (\ref{eq:cRE}), gives
\begin{align}
E = 2 \langle V_T\rangle - \frac{C_a}{8\pi ma} + \frac{R_sC_R}{8\pi m}.  \label{eq:virial}
\end{align}

\subsection{Energy functional}\label{4.4}

We derive the energy functional using the equation of motion for the molecular fields~\cite{Cui20162},
\begin{align}\label{eq:Energy Relation}
E &= \langle H \rangle \nonumber \\
&= \sum_{\sigma=\uparrow,\downarrow}\int\frac{d^3{\bf k}}{(2\pi)^3}\frac{k^{2}}{2m} \left[n_{\sigma}({\bf k}) - \frac{C_{a}}{k^4}\right] + \frac{C_{a}}{4\pi ma} + \frac{R_sC_R}{4\pi m} \nonumber \\
& + \frac{R_s(C_{Q2}+C_{R})}{4\pi m} \left\{ 1 + \frac{\frac{|\Omega_{1}|^{2}}{4}\left[1 + \frac{|\Omega_{2}|^{2}}{4\Delta_{2}^{2}({\bf Q})}\right]}{\left[\Delta_{{\rm e}}({\bf Q}) - i\frac{\gamma_{\rm e}}{2} - \frac{|\Omega_{2}|^{2}}{4\Delta_{2}({\bf Q})} \right]^2 }\right\}^{-1},
\end{align}
where $H$ is the Hamiltonian which is given in the next section. The result above shows that the Raman dressing introduces an extra term into the energy functional, which is related to the molecular energy shift due to the dressing lasers.
Notice that an additional term $\langle V_T\rangle$ should be added to Eq.~(\ref{eq:Energy Relation}) when the system is in the presence of a trapping potential $V_T$.

\section{Contacts in the superfluid state}\label{5}

\begin{figure*}
\includegraphics[width=6.7cm]{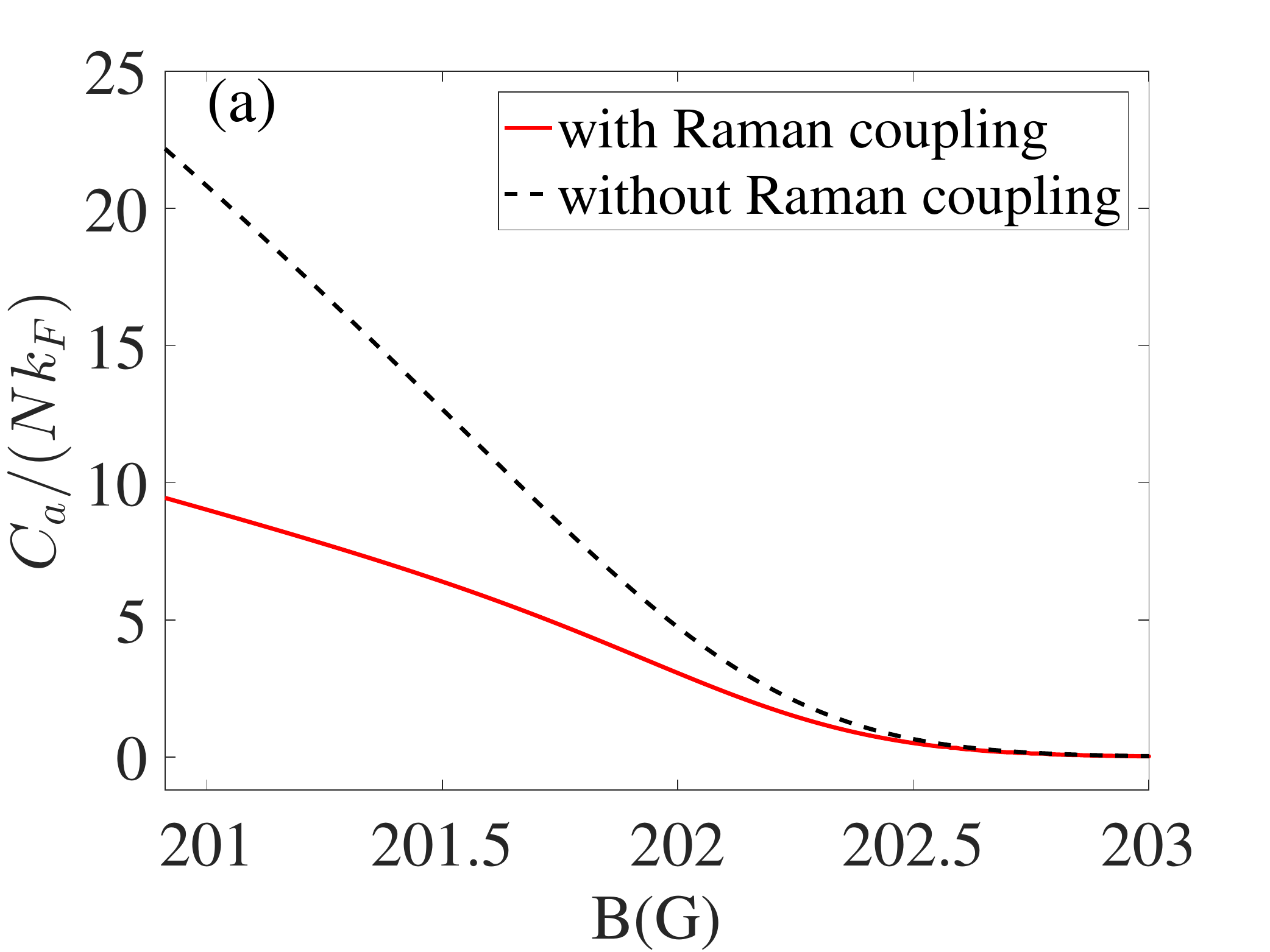}\includegraphics[width=6.7cm]{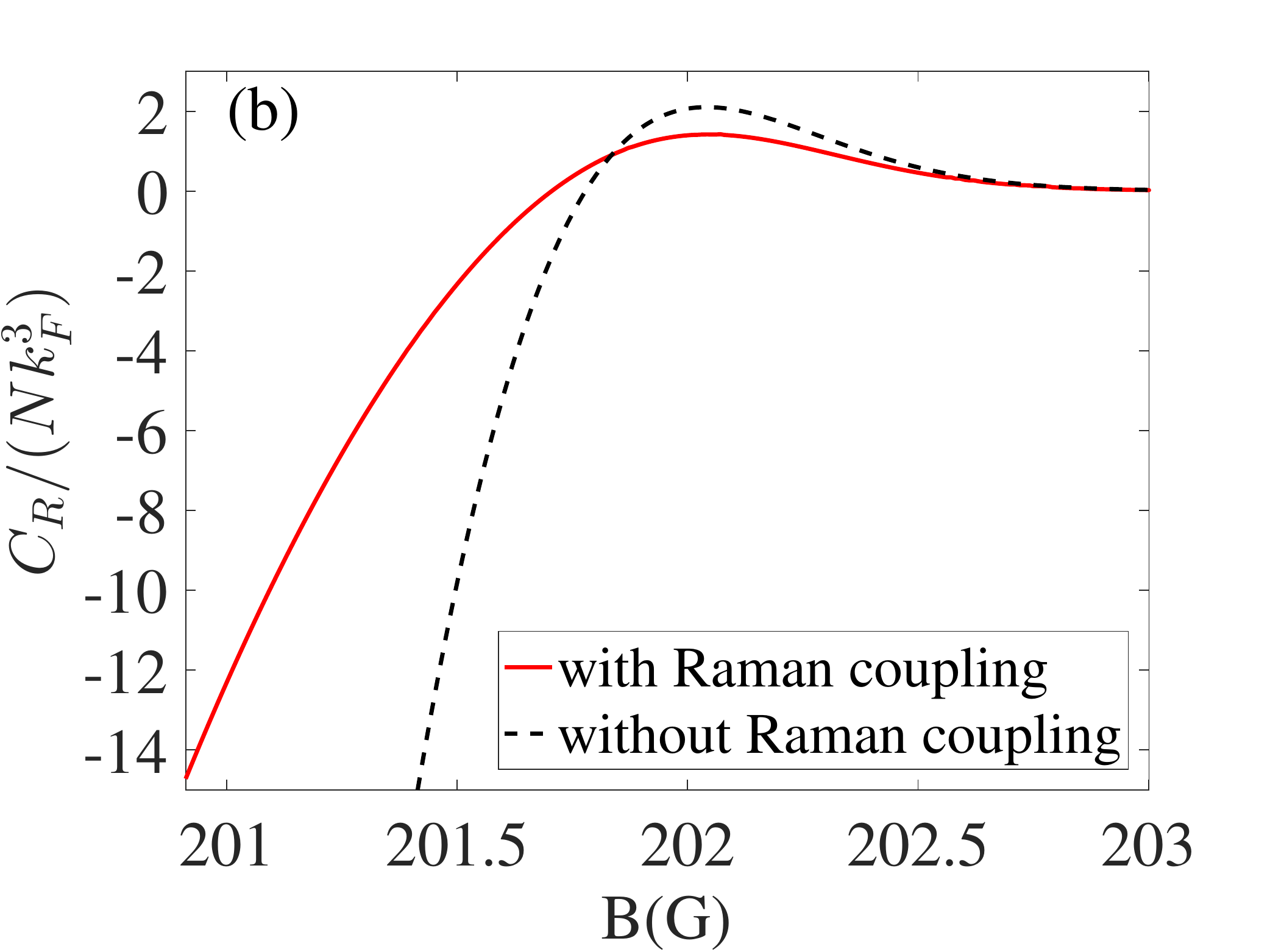}
\includegraphics[width=6.7cm]{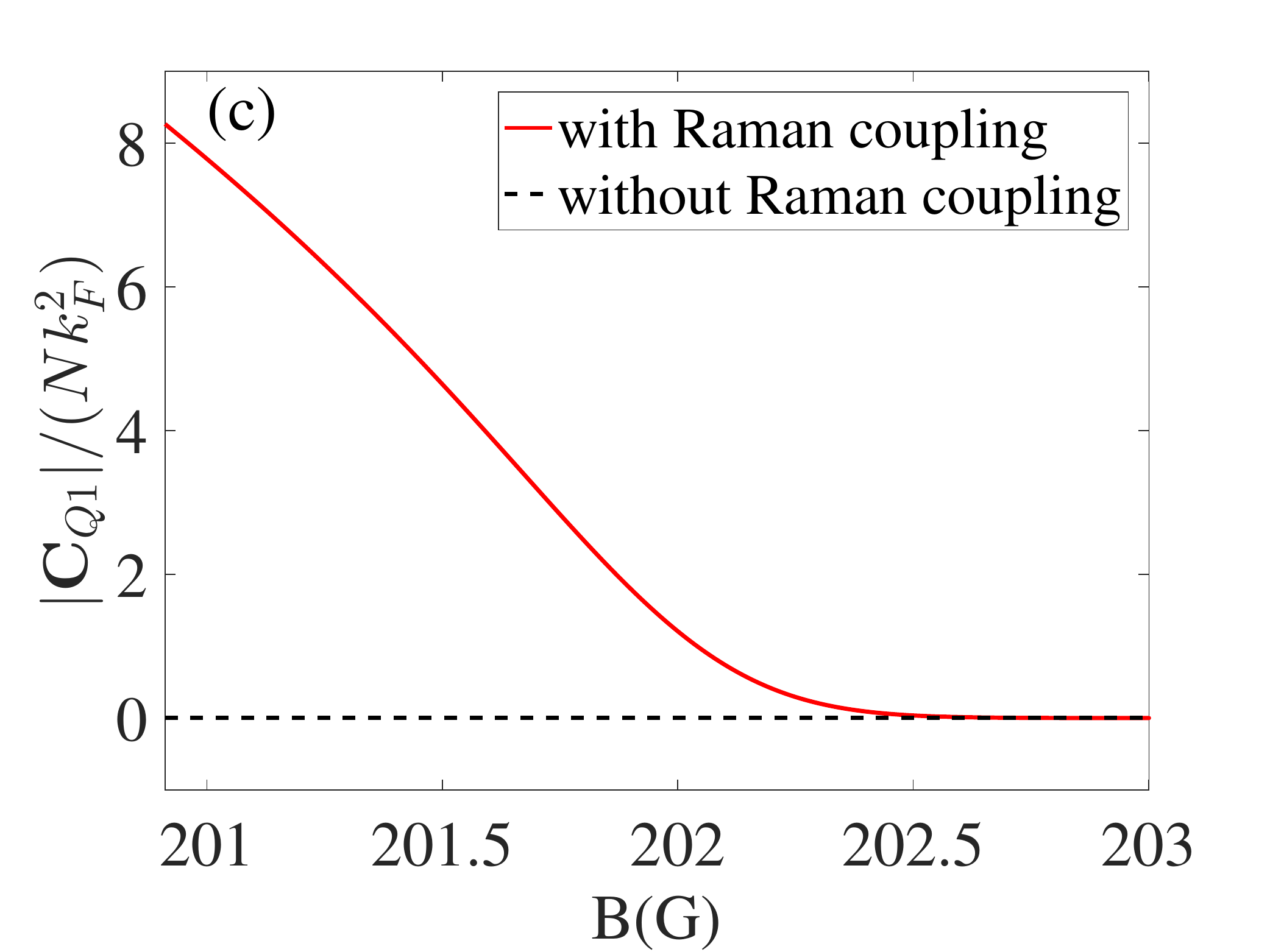}\includegraphics[width=6.7cm]{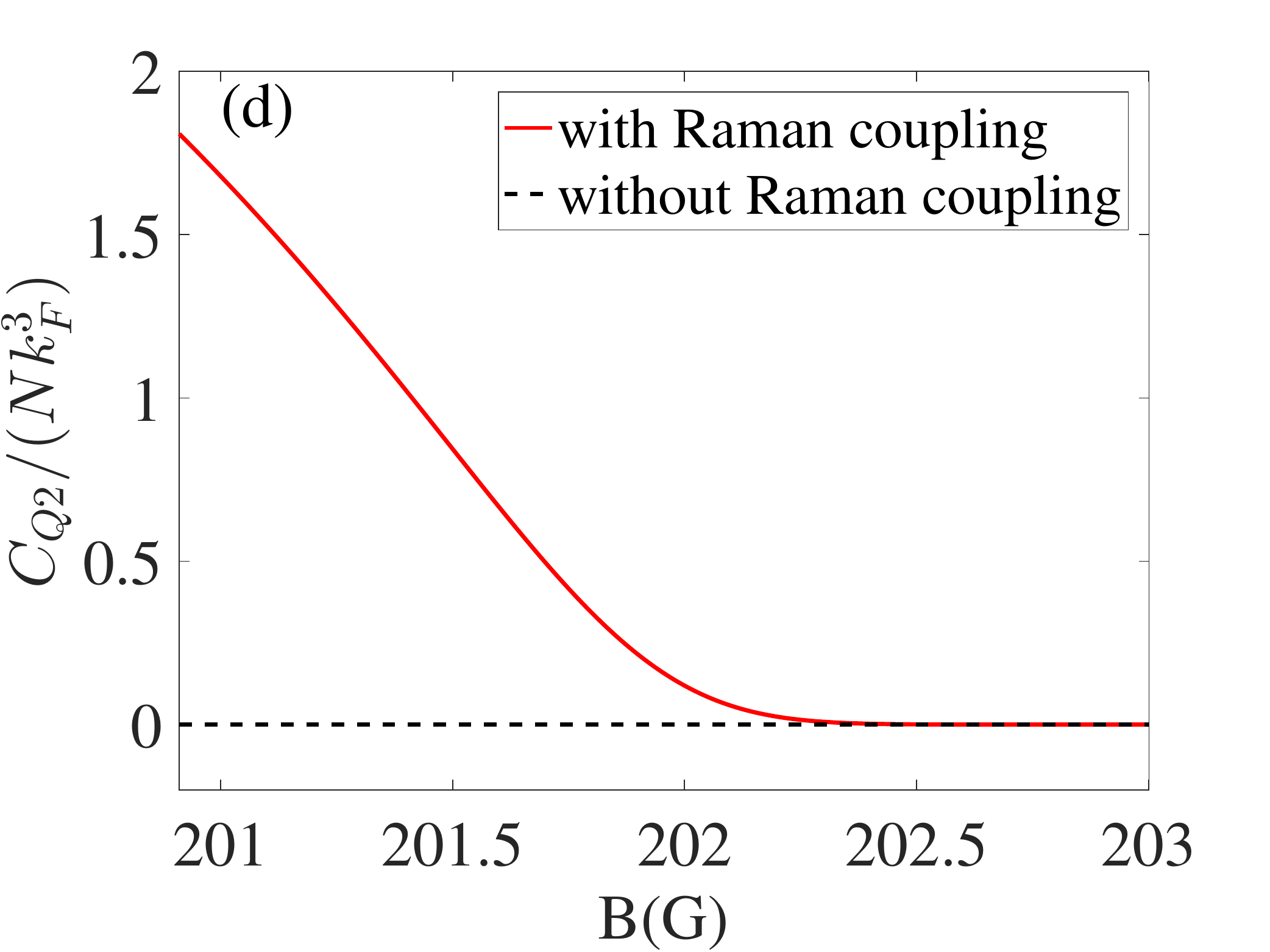}
\includegraphics[width=6.7cm]{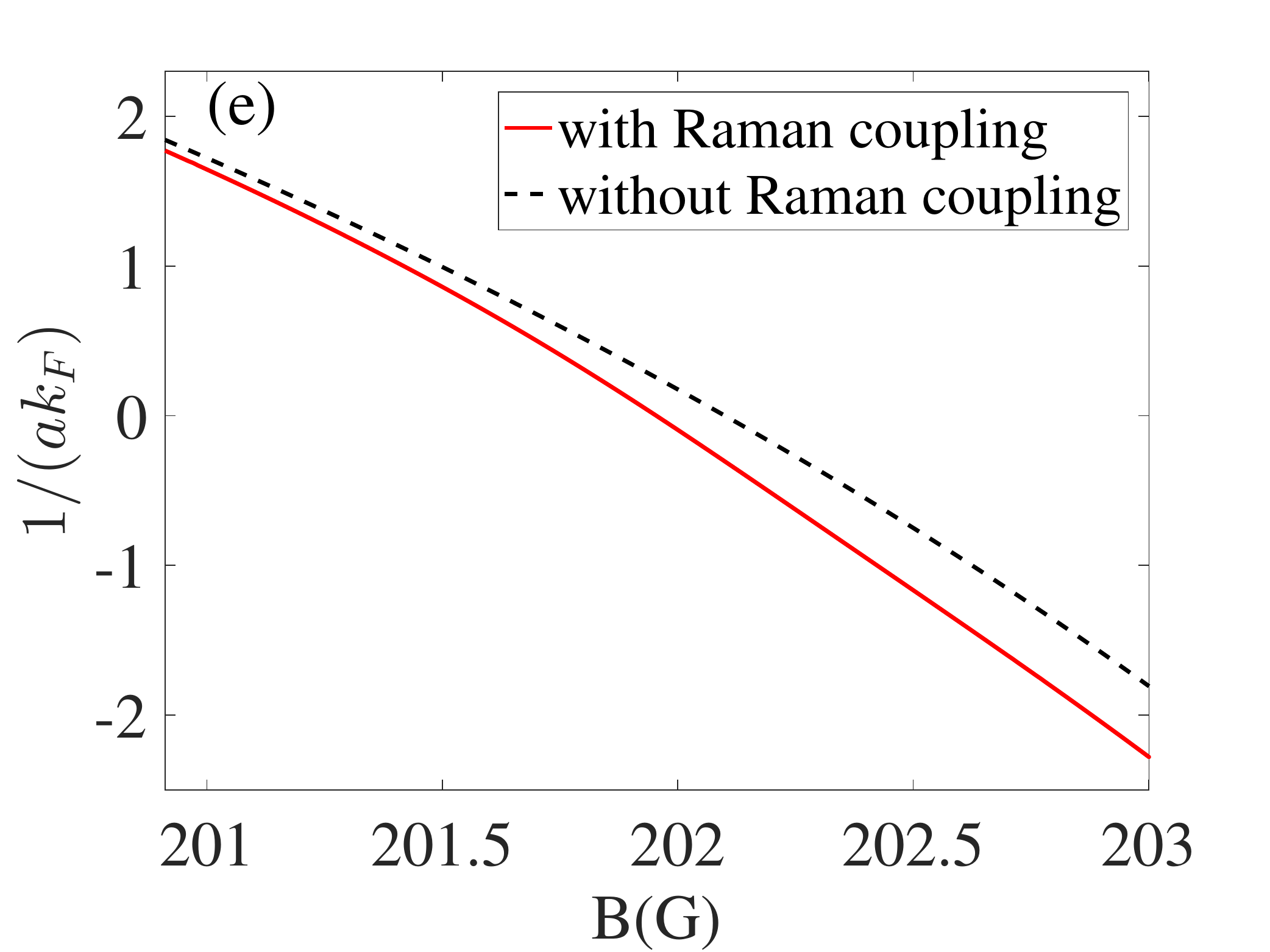}\includegraphics[width=6.7cm]{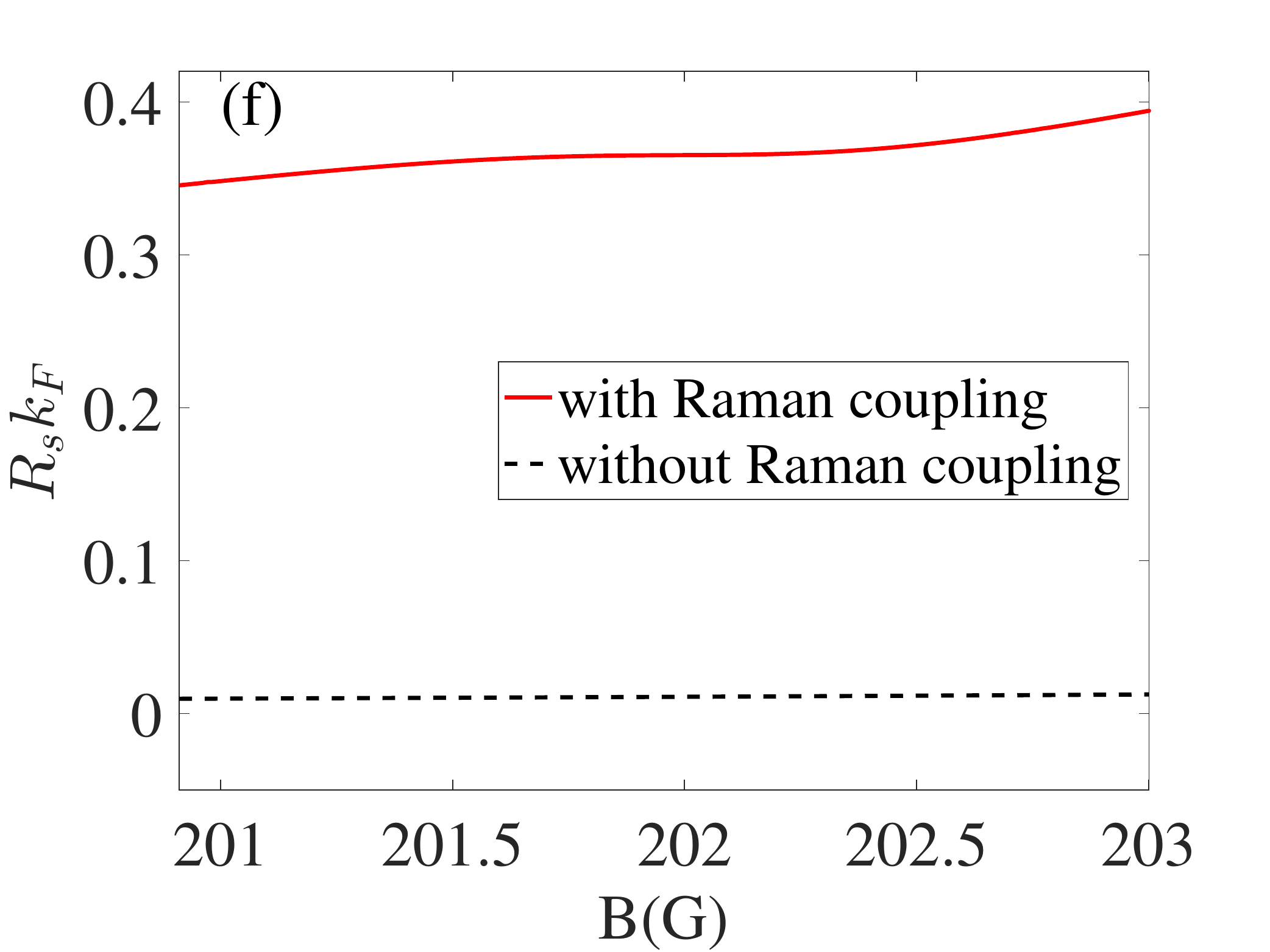}
\caption{(Color online) (a)-(d) Contacts in the superfluid state as functions of the magnetic field for the wide FR of $^{40}$K near $202$ G. (e), (f) Scattering length and range parameter versus the magnetic field, respectively. The red solid lines are calculated under the Raman dressing and the black dashed lines are calculated in the absence of the Raman dressing. Parameters used for the plots are given in the main text.} \label{fig:ContactsFF}
\end{figure*}

To have a better idea of the behavior of contacts under the Raman dressing, we numerically evaluate the contacts using a concrete example.
We consider the zero-temperature superfluid pairing state in the presence of the Raman-induced CoM-momentum-dependent interactions.
The Hamiltonian can be written as
\begin{eqnarray}
H - \mu N &=& H_{{\rm A}} + H_{{\rm M}} + H_{{\rm AM}}, \label{eq:H}\\
H_{{\rm A}} &=& \sum_{{\bf q},\sigma=\uparrow,\downarrow} a_{{\bf q},\sigma}^{\dagger} \left(\frac{k^{2}}{2m} - \mu \right) a_{{\bf q},\sigma}, \label{eq:HA} \\
H_{{\rm M}} &=& \sum_{{\bf Q}} \Phi^{\dagger}_{\bf Q}X(\mu,{\bf Q})\Phi_{\bf Q}, \label{eq:HM}\\
H_{{\rm AM}} &=& \frac{g_{0}}{\sqrt{V}} \sum_{{\bf Q},{\bf q}} \left( b_{{\bf Q},1}^{\dagger} a_{\frac{{\bf Q}}{2}+{\bf q},\uparrow} a_{\frac{{\bf Q}}{2}-{\bf q},\downarrow} \right. \nonumber\\
&&\left. +  a_{\frac{{\bf Q}}{2}-{\bf q},\downarrow}^{\dagger}a_{\frac{{\bf Q}}{2}+{\bf q},\uparrow}^{\dagger}b_{{\bf Q},1}\right) \label{eq:HAM}
\end{eqnarray}
where the matrix $X(\mu,{\bf Q})$ is given by
\begin{eqnarray}
X(\mu,{\bf Q})=\left(\begin{array}{ccc}
-I_{1}(2\mu,{\bf Q}) & 0 & \Omega_{1}^{*}/2\\
0 & -I_{2}(2\mu,{\bf Q}) & \Omega_{2}^{*}/2\\
\Omega_{1}/2 & \Omega_{2}/2 & -I_{{\rm e}}(2\mu,{\bf Q})
\end{array}\right). \nonumber\\
\end{eqnarray}

We assume that the Raman lasers are counterpropagating along the $z$ axis, such that the Raman lasers explicitly break the symmetry of the Fermi surface. The pairing state thus becomes Fulde-Ferrell like, and acquires a finite CoM momentum along the $z$ direction~\cite{He2017}. We then drop the summation over ${\bf Q}$ in the thermodynamic potential and equate ${\bf Q}$ with $Q_{z}{\bf e}_{z}$.
Adopting the mean-field approximation, we define $B_{{\bf Q},\alpha} \equiv \langle b_{{\bf Q},\alpha} \rangle$ ($\alpha=1,2,{\rm e})$ and $\Delta_{{\bf Q}} \equiv \frac{g_{0}}{\sqrt{V}} \langle b_{{\bf Q},1} \rangle$. We write the effective mean-field Hamiltonian as
\begin{align}
&~~H_{m} - \mu N  \nonumber\\
&= \sum_{{\bf q}}\Psi_{\bf Q}^{\dagger}({\bf q}) Y_{\bf q}(\mu,{\bf Q}) \Psi_{\bf Q}({\bf q}) \nonumber\\
&~~ + \sum_{{\bf q}} \left[ \frac{({\bf Q}/2-{\bf q})^{2}}{2m} - \mu \right]
 + B_{\bf Q}^{\dagger} X(\mu,{\bf Q}) B_{\bf Q} , \label{eq:H-mean-field}
\end{align}
where $\Psi_{\bf Q}({\bf q}) = ( a_{\frac{{\bf Q}}{2}+{\bf q},\uparrow}, a_{\frac{{\bf Q}}{2}-{\bf q},\downarrow}^{\dagger} )^{{\rm T}}$ and $B_{\bf Q}=(B_{{\bf Q},1},\ B_{{\bf Q},2},\ B_{{\bf Q},{\rm e}})^{{\rm T}}$. The matrix $Y_{\bf q}(\mu,{\bf Q})$ is given by
\begin{align}
Y_{\bf q}(\mu,{\bf Q})=\left(\begin{array}{cc}
\frac{({\bf Q}/2+{\bf q})^{2}}{2m} - \mu & \Delta_{{\bf Q}}  \\
\Delta_{{\bf Q}}^{*} & -\left[ \frac{({\bf Q}/2-{\bf q})^{2}}{2m} - \mu \right]
\end{array}\right).
\end{align}

Following the standard mean-field approach, we derive the zero-temperature thermodynamic potential,
\begin{align}
\Omega &= -T \ln \text{Tr} e^{-(H_{m} - \mu N)/T} |_{T\rightarrow0} \nonumber\\
&= \sum_{{\bf q},s=\pm} E_{{\bf q},{\bf Q}}^{(s)} \Theta(-E_{{\bf q},{\bf Q}}^{(s)})\nonumber\\
&~~ + \sum_{{\bf q}} \left( \xi_{{\bf q},{\bf Q}} - E_{{\bf q},{\bf Q}} \right) + B_{\bf Q}^{\dagger} X(\mu,{\bf Q}) B_{\bf Q}, \label{eq:omega-mean-field1}
\end{align}
where $\Theta(x)$ is the Heaviside step function, and
$E_{{\bf q},{\bf Q}}^{(\pm)} = E_{{\bf q},{\bf Q}} \pm {\bf q}\cdot{\bf Q}/(2m)$ with $E_{{\bf q},{\bf Q}} = \sqrt{\xi_{{\bf q},{\bf Q}}^2 + |\Delta_{\bf Q}|^2 }$ and $\xi_{{\bf q},{\bf Q}} = q^2/(2m) + Q^2/(8m) - \mu$.


Further, momentum distribution of the spin-$\sigma$ species is
\begin{align}
n_{\sigma}({\bf q})& = \frac{1}{2}\left(1 + \frac{\xi_{{\bf q},{\bf Q},-}}{E_{{\bf q},{\bf Q},-}}\right) \Theta\left(-E_{{\bf q},{\bf Q},-}^{(+)}\right)\nonumber\\
&~~+ \frac{1}{2}\left(1 - \frac{\xi_{{\bf q},{\bf Q},-}}{E_{{\bf q},{\bf Q},-}}\right) \Theta\left(E_{{\bf q},{\bf Q},-}^{(-)}\right), \label{eqn:nq}
\end{align}
where $\xi_{{\bf q},{\bf Q},-}=({\bf q}-{\bf Q}/2)^2/(2m) + Q^2/(8m) - \mu$, $E_{{\bf q},{\bf Q},-}=\sqrt{\xi_{{\bf q},{\bf Q},-}^2 + \Delta_{{\bf Q}}^2}$, and $E_{{\bf q},{\bf Q},-}^{(\pm)} = E_{{\bf q},{\bf Q},-} \pm ({\bf q}-{\bf Q}/2)\cdot{\bf Q}/(2m)$.
Expanding Eq.~(\ref{eqn:nq}) in the large-$q$ limit ($n^{1/3}\ll q\ll 1/r_0$) and matching with Eq.~(\ref{eq:tail}), we immediately have
\begin{align}
&C_a = m^2\Delta_{\bf Q}^2 V,  \label{eq:CaFF} \\
&C_R = \left(2\mu-\frac{Q^2}{4m}\right) m^3\Delta_{\bf Q}^2 V,  \label{eq:CRFF} \\
&{\bf C}_{Q1} = {\bf Q} m^2\Delta_{\bf Q}^2 V,  \label{eq:CQ1FF} \\
&C_{Q2} = \frac{Q^2}{4m} m^3\Delta_{\bf Q}^2 V.  \label{eq:CQ2FF}
\end{align}

To numerically evaluate these contacts, we solve for $Q_z$, $\Delta_{\bf Q}$, and $\mu$ of the ground state from the set of equations $\partial\Omega/\partial B_{{\bf Q},{\rm e}}=0$, $\partial\Omega/\partial B_{{\bf Q},2}=0$, $\partial\Omega/\partial\Delta_{\bf Q}=0$, $\partial\Omega/\partial Q_{z}=0$, together with the number equation $n=-(1/V)\partial\Omega/\partial\mu$. For the numerical calculations, we use the parameters of $^{40}$K atoms. For concreteness, we consider $^{40}$K atoms in the lowest two hyperfine states $|\uparrow\rangle\equiv|F=9/2,m_{F}=-9/2\rangle$ and $|\downarrow\rangle\equiv|F=9/2,m_{F}=-7/2\rangle$. The parameters for the FR near $202$ G are $B_{0}=202.10$ G, $\Delta_{B}=8$ G, $a_{{\rm bg}}=174a_{0}$ with $a_{0}$
being the Bohr radius, and $\delta\mu=1.68\mu_{B}$ with the Bohr magneton $\mu_{B}$~\cite{exp2002s,exp2004s,review2010}. Here we take the atom density $n=1.50\times10^{13}~{\rm  cm}^{-3}$~\cite{exp2004s}.
We consider the typical values $\delta_{{\rm e}}=-2\pi\times500$ MHz, $\gamma_{{\rm e}}=0$, $\delta_{2}=0$, $\Omega_{1}=2\pi\times120$ MHz and $\Omega_{2}=2\pi\times20$ MHz~\cite{Fu2013}. We assume the two laser beams to be counterpropagating along the $z$ axis,
and ${\bf k}_{1}=k_{r}{\bf e}_{z}=-{\bf k}_{2}$ with the recoil momentum $k_{r} \simeq 1.07k_F$ with the two optical frequencies $\omega_{1}\approx\omega_{2}=2\pi\times3.9\times10^{14}$ Hz~\cite{Fu2013,Zhangexp2017,Thomasexp2016}.

Figure~\ref{fig:ContactsFF} shows the contacts of $^{40}$K atoms as functions of the magnetic field in a superfluid state.
The red solid lines are the contacts in the presence of the Raman dressing, and the black dashed lines denote the contacts in the absence of the laser dressing. Apparently, the Raman dressing has significant impact on the contacts. Most prominently, as the Raman dressing gives rise to a Fulde-Ferrell pairing state, ${\bf C}_{Q1}$ and $C_{Q2}$ acquire finite values only in the presence of the laser dressing.

\section{Summary}\label{6}

We have shown that in a three-dimensional Fermi gas with Raman-dressed FR, the high-momentum tail of the density distribution can be characterized by four CoM-momentum-dependent contacts. These contacts determine the leading $q^{-4}$ and the subleading $q^{-5}$ and $q^{-6}$ tails in the distribution and appear in various universal relations. Among the four contacts, we demonstrate that two are related to the scattering length and the range parameter, respectively. The remaining two contacts are related to the CoM motion of closed-channel molecules. We find that the $q^{-5}$ tail and part of the $q^{-6}$ tail of the momentum distribution show anisotropic behaviors. We derive the universal relations, and numerically estimate contacts for the zero-temperature superfluid state under the CoM-dependent interaction. Our results shed light on the behaviors of high-momentum distribution and the universal relations in cold-atom gases with dressed FR, and can be readily checked experimentally.

\section*{Acknowledgements}

We thank Peng Zhang, Lianyi He, Xiaoling Cui, Guo-Zhu Liu, Ming Gong, Lijun Yang, and Jing-Bo Wang for useful discussions. This work is supported by the National Natural Science Foundation of China (Grants No. 60921091, No. 11374283, No. 11522545, No. 11404106) and
the National Key R\&D Program of China (Grants No. 2016YFA0301700, No. 2017YFA0304800). W.Y. acknowledges support from the ``Strategic Priority Research Program(B)'' of the Chinese Academy of Sciences, Grant No. XDB01030200.
F.Q. acknowledges support from the Project funded by the China Postdoctoral Science Foundation (Grant No. 2016M602011).

\end{CJK*}
\end{document}